\documentclass[onecolumn,draftclsnofoot]{IEEEtran}%
\usepackage{cite}%

\ifCLASSINFOpdf
\else
\fi
\usepackage{amssymb}
\usepackage[cmex10]{amsmath}
\usepackage{amsthm}

\newtheorem{theorem}{Theorem}
\newtheorem{definition}[theorem]{Definition}

\begin{document}
%
\title{
Generalization of Rashmi-Shah-Kumar Minimum-Storage-Regenerating Codes
}
%
%
%

\author{Masazumi~Kurihara  and  Hidenori~Kuwakado
\thanks{M. Kurihara is with the 
Graduate School of Informatics and Engineering, 
University of Electro-Communications, 
1-5-1 Chofugaoka Chofu Tokyo 182-8585 Japan, 
e-mail: kurihara@uec.ac.jp.}
\thanks{H. Kuwakado is with 
the Faculty of Informatics, Kansai University, 
2-1-1 Ryozenji-cho Takatsuki-shi Osaka, 569-1095 Japan, 
e-mail: kuwakado@kansai-u.ac.jp.
}

}

\maketitle

\begin{abstract}
In this paper, we propose a generalized version of 
the Rashmi-Shah-Kumar Minimum-Storage-Regenerating(RSK-MSR) codes 
based on the product-matrix framework. 
For any $(n,k,d)$ such that $d \geq 2k-2$ and $d \leq n-1$, 
we can directly construct an $(n,k,d)$ MSR code 
without 
constructing a larger MSR code and shortening of the larger MSR code.
As a result, the size of a finite field over which the proposed code is defined is smaller than or equal to the size of a finite field  over which the RSK-MSR code is defined.
In addition, 
the $\{\ell,\ell'\}$ secure codes based on the generalized RSK-MSR codes 
can be obtained by applying
the construction method of $\{\ell,\ell'\}$ secure codes 
proposed by Shah, Rashmi and Kumar. 
Furthermore, the message matrix of the $(n,k,d)$ generalized RSK-MSR code 
is derived from that of the RSK-MSR code 
by using the construction method of the $\{\ell=k,\ell'=0\}$ secure code.  
\end{abstract}

\begin{IEEEkeywords}
Distributed storage, 
regenerating codes, 
Minimum-Storage-Regenerating codes(MSR codes), 
Rashmi-Shah-Kumar MSR codes(RSK-MSR codes), 
generalized RSK-MSR codes,  
secure regenerating codes. 
\end{IEEEkeywords}

%
\IEEEpeerreviewmaketitle

\section{Introduction}

Dimakis, Godfrey, Wu, Wainwright and Ramchandran 
introduced a concept of regenerating codes 
into distributed storage systems 
\cite{dimakis}. 
Under the concept, a regenerating code with six parameters $(n,k,d,\alpha,\beta,B)$  
has two properties of reconstruction and regeneration as follows. 
A message consists of $B$ message symbols over a finite field 
$\mathbb{F}_{q}$ with $q$ elements. 
The message is encoded to $n$ shares in such a way that 
the message can be reconstructed from any $k$ shares, 
and the $n$ shares are stored across $n$ storage nodes in 
a distributed storage system. 
Each share consists of $\alpha$ symbols over $\mathbb{F}_{q}$, i.e., 
we assume that the storage capacity of each node is $\alpha$ symbols. 
A data collector is permitted to connect to any $k$ active nodes to reconstruct the message, 
and downloads the $k$ shares from the $k$ nodes. 
Then, the data collector can reconstruct the message from the $k$ shares. 
Furthermore, 
a failed node is permitted to connect to any $d$ active nodes, 
which are called helper nodes, to regenerate the share as was stored in itself, 
and downloads data consisting of $\beta$ symbols over $\mathbb{F}_{q}$ from each helper node. 
As a result, the failed node obtains the downloaded data of total amount $d\beta$, 
and 
can regenerate the share from the downloaded data, 
i.e., the failed node can be repaired. 
The total amount $d\beta$ of the downloaded data for repair is called the repair-bandwidth. 

Dimakis et al. \cite{dimakis} also showed 
that there is a fundamental tradeoff between storage $\alpha$ and repair-bandwidth $d\beta$. 
In the optimal tradeoff, 
there are Minimum Storage Regenerating(MSR) codes and Minimum Bandwidth Regenerating(MBR) codes 
as optimal regenerating codes.  
In particular, an $(n,k,d)$ MSR code with $(\alpha,\beta,B)$ satisfies the following optimal condition. 
\begin{align}
\label{eq:1}
(\alpha,\beta)=\left(\frac{B}{k},\frac{B}{k(d-k+1)} \right).
\end{align}
When the value of $\beta$ is equal to one, i.e., $\beta=1$, 
the parameters $\alpha$ and $B$ are uniquely decided as follows: 
\begin{align}
\label{eq:2}
\alpha &= d-k+1, \\
\label{eq:3}
B &= k \alpha  = k(d-k+1).
\end{align}

In this paper, we focus on a construction of MSR codes. 
The constructions of MSR codes are proposed in 
\cite{rashmi,suh,shah}. 

Rashmi, Shah and Kumar proposed 
a construction of $(n,k,d=2k-2)$ MSR codes based on product-matrix framework, 
where $d \leq n-1$ and $\beta=1$  
\cite[Sec. V]{rashmi}
(see Table \ref{table:nkd-1}). 
Furthermore, 
for any $(n,k,d)$ such that $d \geq 2k-2$ and $d \leq n-1$,  
an $(n,k,d)$ MSR code with $\beta=1$ is derived from 
a larger $(n'=n+(d-2k+2), k'=k+(d-2k+2), d'=d+(d-2k+2))$ MSR code   
by using a procedure of shortening of the larger MSR code, 
where $d'=2k'-2$ 
\cite[Sec. V-C, Corollary 8]{shah}
(see Table \ref{table:nkd-2}).  
The larger $(n',k',d')$ MSR code can be constructed by \cite[Sec. V]{rashmi}. 
In particular, in this paper,  
the MSR code is called a Rashmi-Shah-Kumar(RSK) MSR code, i.e., RSK-MSR code.   

Suh and Ramchandran\cite{suh} and 
Shah, Rashmi, Kumar and Ramchandran\cite{shah} 
proposed MSR codes based on interference alignment techniques. 
Suh and Ramchandran proposed  
a construction of $(2k,k,d=2k-1)$ MSR codes, 
which are called Exact-Repair MDS codes, where $q \geq 2k$  and $\beta=1$  
\cite[Sec. VI-A, Theorem 2]{suh}
(see Table \ref{table:nkd-1}). 
Moreover,   
for any $(n,k,d)$ such that $n \geq 2k$ and $d \geq 2k-1$, 
an $(n,k,d)$ MSR code  with $\beta=1$ is derived from 
a larger $(n'=2(n-k), k'=n-k, d'=2(n-k)-1)$ MSR code  
by using procedures of removing nodes and pruning equations, 
where $d'=2k'-1$ \cite[Sec. VI-B, Theorem 3]{suh}
(see Table \ref{table:nkd-2}). 
The larger $(n',k',d')$ MSR code can be constructed by \cite[Sec. VI-A, Theorem 2]{suh}. 

On the other hand, 
Shah et al. proposed 
a construction of $(2k,k,d=2k-1)$ MSR codes which are called MISER
codes\footnote{
Short for MDS, Interference-aligning, Systematic, Exact-Regenerating codes
}, 
where $\beta=1$ \cite[V-B]{shah}
(see Table \ref{table:nkd-1}). 
Furthermore, 
for any $(n,k,d)$ such that $n \geq 2k$ and $d =n-1$, 
an $(n,k,d)$ MSR code  with $\beta=1$ is derived from 
a larger $(n'=n+(n-2k), k'=k+(n-2k), d'=d+(n-2k))$ MSR code 
by using a procedure of shortening of the larger MSR codes, 
where $d'=2k'-1$ \cite[Sec. V-C]{shah}
(see Table \ref{table:nkd-2})\footnote{
Under the additional constraint in regeneration, 
Shah et al. showed the extension of the MISER code to the case $2k-1\leq d \leq n-1$  
\cite[Sec. V-D]{shah}.
}. 
The $(n',k',d')$ MSR code can be constructed by \cite[Sec. V-B]{shah}. 
The relation between codes of \cite{shah} and \cite{suh} is written in 
\cite[Sec. I-D and V-G]{shah} and \cite[Sec. II-C]{suh}. 

\begin{table*}[tbh]
\caption{
\rm Relations of parameters and conditions 
of $(n,k,d)$ MSR codes with $\beta=1$ over $\mathbb{F}_{q}$ in \cite{rashmi,suh,shah}.
 }
\label{table:nkd-1}
\begin{center}
{\footnotesize
\begin{tabular}{l|ll}
\hline
 & Parameters $(n,k,d)$ & Conditions \\
\hline
Rashmi {et al}. & $(n,k,2k-2)$ &  $d=2k-2$, $d \leq  n-1$,  \\
\cite[Sec. V]{rashmi} & & $q \geq n(d-k+1)$ \\
\hline
Suh {et al}. & $(2k,k,2k-1)$ & $d=2k-1$, $d=n-1$,  \\
\cite[Sec. VI-A]{suh} & &  $q \geq 2k$ \\
\hline
Shah {et al}. & $(2k,k,2k-1)$ & $d=2k-1$, $d=n-1$, \\
\cite[Sec. V-B]{shah}
 & &  $q \geq 2k$ ${}^{\dag}$
 \\ %
\hline
Presented codes & $(n,k,d)$ & $d \geq 2k-2$, $d \leq n-1$,  \\
in this paper & &  $q \geq n(d-k+1)$  \\
\hline
\end{tabular}\\
${}^{\dag}${
The minimum field size is derived from \cite[Eq.(37)]{shah} as follows: 
$q \geq \alpha+n-k=n=2k$. 

}
}
\end{center}
\end{table*}

\begin{table*}[tbh]
\caption{
\rm Relations of parameters and conditions 
of the larger $(n',k',d')$ MSR codes 
to construct the target $(n,k,d)$ MSR codes with $\beta=1$ 
over $\mathbb{F}_{q}$ 
in \cite{rashmi,suh,shah}, 
where $i_{1}=d-2k+2$ and $i_{2}=n-2k$.  
}
\label{table:nkd-2}
\begin{center}
{\footnotesize
\begin{tabular}{l|ll}
\hline
 & Parameters $(n',k',d')$ & Conditions \\
\hline
Rashmi {et al}.
 & $( n+i_{1},k+i_{1}, d+i_{1} )$ 
 & $d \geq 2k-2$, $d \leq n-1$,\\
\cite[Sec. V-C]{rashmi}  &
 & $q\geq (n+d-2k+2)(d-k+1)$ \\
\hline
Suh {et al}.
 & $( 2(n-k),n-k,2(n-k)-1 )$ 
 & $d \geq 2k-1$,  $n\geq 2k$, \\  
\cite[Sec. VI-B]{suh}  &
 & $q \geq 2(n-k)$\\
\hline
Shah {et al}.
 & $( n+i_{2},k+i_{2},d+i_{2} )$ 
 & $n\geq 2k$, $d =n-1$,   \\
\cite[Sec. V-C]{shah}  &
 & $q \geq 2(n-k)$\\
\hline
\end{tabular}
}
\end{center}
\end{table*}

In this paper, we propose a generalized version of RSK-MSR codes, which are based on product-matrix framework,  
proposed by Rashmi et al.\cite{rashmi}. 
For any $(n,k,d)$ such that $d \geq 2k-2$ and $d \leq n-1$, 
we can directly construct an $(n,k,d)$ MSR code with $\beta=1$ 
without using processes of constructing a larger MSR code and shortening of the larger MSR code
(see Table \ref{table:nkd-1}). 
We will call the presented code in this paper 
a generalized RSK-MSR code.

This paper is organized as follows: 
In section 
\ref{sec:construction}, 
a construction of generalized RSK-MSR codes is proposed. 
Furthermore, 
the reconstruction and regeneration of the code are described. 
In section 
\ref{sec:example}, 
examples of construction, reconstruction and regeneration are 
given.  
In section 
\ref{sec:securecode}, 
relations between generalized RSK-MSR codes and 
$\{\ell,\ell'\}$ secure codes are described. 
Finally, a conclusion is given in section 
\ref{sec:conclusion}.

\section{Generalized RSK-MSR Codes}
\label{sec:construction}

In this section, 
we propose a construction of $(n,k,d)$ MSR code with $\beta=1$ 
over $\mathbb{F}_{q}$ 
for any $(n,k,d)$ such that $d \geq 2k-2$ and $d \leq n-1$. 
Then, 
the remaining parameters $\alpha$ and $B$ are uniquely 
determined from $k,d$ and $\beta$ as follows: 
$\alpha=d-k+1$ and $B=k\alpha=k(d-k+1)$. 


We assume that there are $n$ storage nodes such as node $i$, $1 \leq i \leq n$, 
in a distributed storage system. 
The storage capacity of each node is $\alpha$ symbols over $\mathbb{F}_{q}$.  
Furthermore,
for each $i \in \{1,\dots,n \}$,  
assign an unique public symbol $x_{i}$ in $\mathbb{F}_{q}$ to 
node $i$ in such a way that the following conditions are satisfied. 
\begin{enumerate}
\item For any $i \in \{ 1,\dots,n \}$, $x_{i} \neq 0$, 
\item For any $i,j \in \{ 1,\cdots,n \}$, $x_{i}^{\alpha} \neq x_{j}^{\alpha}$ if $i \neq j$.  
\end{enumerate}

From the condition of parameters $q,n$ and $\alpha$ written in \cite[Sec. V]{rashmi}, 
if the condition, $q \geq n \alpha$, is true, then 
there are 
at least $n$ elements $x_{1},\dots,x_{n}$ in $\mathbb{F}_{q}$ 
such that $x_{i}^{\alpha} \neq x_{j}^{\alpha}$ if $i\neq j$. 
The condition is the sufficient condition of the existence of such $n$ elements, 
but it is not the necessary condition of that
(see Example in Section \ref{sec:example}). 

In general, 
we can construct an $(n,k,d)$ generalized RSK-MSR code over $\mathbb{F}_{q}$, 
where $q \geq n(d-k+1)$. 
On the other hand, 
in the construction method proposed by Rashmi et al.\cite{rashmi}, 
the target $(n,k,d)$ RSK-MSR code can be constructed 
by shortening a larger 
$( n'=n+d-2k+2, k'=d-k+2, d'=2(d-k+1) )$ RSK-MSR code 
over $\mathbb{F}_{q_{0}}$,   
where $q_{0} \geq n'(d'-k'+1)=\{n+(d-2k+2)\}(d-k+1)$. 
Thus, the size of the finite field  $\mathbb{F}_{q}$ is smaller than or equal to 
that of the finite field $\mathbb{F}_{q_{0}}$,   
since $d \geq 2k-2$. 

\subsection{Message Matrix $M$}
\label{subsec:message}

Firstly, 
for given parameters $k$ and $d$,  
we define five types of $(d\times(d-k+1))$ message 
matrices $M$ consisting of 
the following sub-matrices.  
Note that  the message matrix is also represented 
as a $(d\times\alpha)$ matrix because $\alpha = d-k+1$. 
\begin{align*} 
\begin{array}{rl}
T_{1}, T_{2}  : & \text{ $((k-1)\times(k-1))$ symmetric matrices,} \\
U_{1}  : & \text{a $((k-1)\times 1)$ matrix, i.e., a column vector of length $k-1$,} \\
U_{2}  : & \text{a $(1\times 1)$ matrix, i.e., a scalar,} \\
V_{1}  : & \text{a $((k-1)\times (d-2k+1))$ matrix,} \\
V_{2}  : & \text{a $( 1 \times (d-2k+1))$ matrix, i.e., a row vector of length $d-2k+1$,} \\
O_{1}  : & \text{a $((d-2k+1) \times (d-2k+1))$ square zero matrix,} \\
O_{2}  : & \text{a $((k-1) \times (d-2k+2))$ zero matrix.} 
\end{array}
\end{align*} 
\begin{enumerate}
\item Type I : 
When $k \geq 2$ and $d=2k-2$, 
the message matrix is defined as
\begin{align} 
M=\left[
\begin{array}{c}
T_{1} \\    
T_{2} 
\end{array}
\right].
\end{align} 
The message matrix of Type I is identical with that of the RSK-MSR code\cite{rashmi}.  
The message matrix is a $(d\times (d-k+1))$ matrix because $d=2k-2$. 
Since 
$T_{1}$ and $T_{2}$ are $((k-1)\times(k-1))$ symmetric matrices, 
each of the two matrices consists of $k(k-1)/2$ distinct entries. 
Thus, the message matrix consists of $B=k(k-1)$ distinct message symbols 
because $B=k(d-k+1)=k(k-1)$ when $d=2k-2$. 

\item Type II : 
When $k \geq 2$ and $d=2k-1$, 
the message matrix is defined as
\begin{align} 
M=\left[
\begin{array}{c|c}
T_{1}     & \multicolumn{1}{c}{U_{1}}  \\
\hline
U_{1}^{t} & \multicolumn{1}{c}{U_{2}}  \\
\hline
T_{2} & \multicolumn{1}{c}{O_{2}}  
\end{array}
\right], 
\end{align} 
where $U_{1}^{t} $ is the transpose of $U_{1}$. 
The message matrix of Type II is a $(d\times (d-k+1))$ matrix because 
$d=2k-1$. 
The sub-matrix
$
\left[
\begin{smallmatrix}
T_{1}     & U_{1} \\
U_{1}^{t} & U_{2} 
\end{smallmatrix}
\right]
$ 
of $M$ is a $(k\times k)$ symmetric matrix and 
the $(k\times 1)$ sub-matrix
$
\left[
\begin{smallmatrix}
U_{1} \\
U_{2} 
\end{smallmatrix}
\right]
$ 
consists of $k$ entries. 
Thus,   
the message matrix consists of $B=k^{2}$ distinct message symbols 
because $B=k(d-k+1)=k^{2}$ when $d=2k-1$. 

\item Type III : 
When $k \geq 2$ and $d\geq 2k$, 
the message matrix is defined as
\begin{align} 
M=\left[
\begin{array}{c|cc}
T_{1}     & \multicolumn{1}{c|}{U_{1}} & V_{1} \\
\hline
U_{1}^{t} & \multicolumn{1}{c|}{U_{2}}     & V_{2} \\
\hline
V_{1}^{t} & \multicolumn{1}{c|}{V_{2}^{t}} & O_{1} \\
\hline
T_{2} & \multicolumn{2}{c}{O_{2}}  
\end{array}
\right].
\end{align} 
The message matrix of Type III is a $(d\times (d-k+1))$ matrix. 
Since the sub-matrix
$
\left[
\begin{smallmatrix}
T_{1}     & U_{1} & V_{1} \\
U_{1}^{t} & U_{2} & V_{2} \\
V_{1}^{t} & V_{2}^{t} & O_{1} 
\end{smallmatrix}
\right]
$ 
of $M$ is a $((d-k+1)\times (d-k+1))$ symmetric matrix and 
the $(k \times (d-2k+2))$ sub-matrix
$
\left[
\begin{smallmatrix}
 U_{1} & V_{1} \\
 U_{2} & V_{2} 
\end{smallmatrix}
\right]
$ consists of $k(d-2k+2)$ entries, 
the message matrix consists of $B=k(k-d+1)$ distinct message symbols. 

\item Type IV : 
When $k =1$ and $d=1$, 
the message matrix is defined as
\begin{align} 
M=\left[
\begin{array}{c}
U_{2}
\end{array}
\right]. 
\end{align} 
The message matrix of Type IV is a $(d\times (d-k+1))$ matrix 
because $d-k+1=d=1$ when $k=1$ and $d=1$.   
Thus, 
the message matrix consists of $B=1$ message symbol 
because $B=k(d-k+1)=d=1$ when $k=1$ and $d=1$. 

\item Type V : 
When $k =1$ and $d\geq 2$, 
the message matrix is defined as
\begin{align} 
M=\left[
\begin{array}{c|c}
U_{2}     & \multicolumn{1}{c}{V_{2}}  \\
\hline
V_{2}^{t} & \multicolumn{1}{c}{O_{1}}  
\end{array}
\right]. 
\end{align} 
The message matrix of Type V is a $(d\times (d-k+1))$ matrix 
because $d-k+1=d$ when $k=1$.  
The message matrix 
is a $(d\times d)$ symmetric matrix 
and  
the sub-matrix $O_{1}$ is a $((d-1)\times(d-1))$ square zero matrix.  
Thus,  
the message matrix consists of $B=d$ distinct message symbols 
because $B=k(d-k+1)=d$ when $k=1$. 
\end{enumerate}
From the above definition, 
these five types of the message matrices for $k$ and $d$ are 
coordinated in Table \ref{table:Mkd}. 
The message matrix of every type is a $(d \times (d-k+1))$ matrix and 
consists of $B$ message symbols. 
\begin{table}[htb]
\caption{
\rm Types of the message matrices $M$ for $k$ and $d$ such that $d \geq 2k-2$. 
 }
\label{table:Mkd}
\begin{center}
{\footnotesize
\begin{tabular}{c||c|c|c}
\hline
$k=1$ & $(d=1)$ & \multicolumn{2}{|c}{$(d\geq 2)$} \\
      & IV & \multicolumn{2}{|c}{V} \\
\hline
\hline
$k\geq 2$ & $(d=2k-2)$ & $(d=2k-1)$ & $(d\geq2k)$  \\
          & I & II & III \\
\hline
\end{tabular}
}
\end{center}
\end{table}
%


\subsection{Encoding and Share}
\label{subsec:encoding}

For each $i$, $1 \leq i \leq n$, 
we define a coding vector $\underline{\rho}_{i}$ associated with node $i$ as follows:
\begin{align} 
\underline{\rho}_{i}
= [ 1,x_{i},x_{i}^{2},\dots,x_{i}^{d-1} ] \in \mathbb{F}^{d},
\end{align} 
where $x_{i}$ is the element assigned to node $i$.  
The message consisting of $B$ message symbols is encoded to $n$ shares by 
the coding vectors $\underline{\rho}_{i}$, $1 \leq i \leq n$, 
and the message matrix $M$.  
For each $i$, $1 \leq i \leq n$, 
the share $\underline{c}_{i}$, which is stored in node $i$, is defined by 
\begin{align} 
\underline{c}_{i}
=[ c_{i,1},\dots,c_{i,\alpha} ]
= \underline{\rho}_{i}M
\in \mathbb{F}^{\alpha}.
\end{align}


\subsection{Reconstruction}
\label{subsec:reconstruction}

In this section, 
we describe a reconstruction of the generalized RSK-MSR code.  
%
Before describing the reconstruction, 
we write the coding vector and the share using the following two sub-vectors.
%
\begin{align} 
{\underline{\omega}_{i}}&=[{ 1,x_{i},\dots,x_{i}^{k-2} }] 
\in \mathbb{F}^{k-1}, \\
{\underline{\theta}_{i}}&=[{ 1, x_{i}, \dots, x_{i}^{\alpha-k-1} }]
\in \mathbb{F}^{\alpha-k}.
\end{align} 
\begin{enumerate}
\item In the case of Type I, the coding vector is represented as
\begin{align*} 
\underline{\rho}_{i}
=
[
\underbrace{ 1,\dots,x_{i}^{k-2} }_{\underline{\omega}_{i}},
\underbrace{ x_{i}^{k-1},  \dots,x_{i}^{d-1} }_{x_{i}^{\alpha}\underline{\omega}_{i}}
],  
\end{align*} 
where $\alpha=k-1$ because $d=2k-2$, 
and then, the components of the share are represented as
\begin{align*} 
[ c_{i,1},\dots,c_{i,\alpha} ]
=
\left[ \underline{\omega}_{i},  
       x_{i}^{\alpha} \underline{\omega}_{i} \right]
\left[
 \begin{array}{c}
T_{1} \\
T_{2}
 \end{array}
\right].
\end{align*} 

\item In the case of Type II, the coding vector is represented as
\begin{align*} 
\underline{\rho}_{i}
=
[
\underbrace{ 1,\dots,x_{i}^{k-2} }_{\underline{\omega}_{i}},
x_{i}^{k-1}, 
\underbrace{ x_{i}^{k},  \dots,x_{i}^{d-1} }_{x_{i}^{\alpha}\underline{\omega}_{i}}
],  
\end{align*} 
where $\alpha=k$ because $d=2k-1$, 
and then, the components of the share are represented as
\begin{align*} 
[ c_{i,1},\dots,c_{i,\alpha-1} ]
&=
\left[ \underline{\omega}_{i} ,    
       x_{i}^{\alpha} \underline{\omega}_{i} \right]
\left[
 \begin{array}{c}
T_{1} \\
T_{2}
 \end{array}
\right]
+
 x_{i}^{k-1} U_{1}^{t}, \\
c_{i,\alpha}
&=
\left[ \underline{\omega}_{i},  x_{i}^{k-1}  \right] 
\left[
 \begin{array}{c}
 U_{1} \\
 U_{2}
 \end{array}
\right].  
\end{align*} 

\item In the case of Type III, the coding vector is represented as
\begin{align*} 
\underline{\rho}_{i}
=
[
\underbrace{ 1,\dots,x_{i}^{k-2} }_{\underline{\omega}_{i}},
x_{i}^{k-1}, 
\underbrace{ x_{i}^{k},  \dots,x_{i}^{\alpha-1} }_{x_{i}^{k}\underline{\theta}_{i}},
\underbrace{ x_{i}^{\alpha}, \dots,x_{i}^{d-1} }_{x_{i}^{\alpha}\underline{\omega}_{i}}
],  
\end{align*} 
where $\alpha \geq k+1$ because $d\geq 2k$, 
and then, the components of the share are represented as
\begin{align*} 
[ c_{i,1},\dots,c_{i,k-1} ]
 &= 
\left[ \underline{\omega}_{i} ,   
       x_{i}^{\alpha} \underline{\omega}_{i}  \right]
\left[
 \begin{array}{c}
T_{1} \\
T_{2}
 \end{array}
\right]
 +
\left[ x_{i}^{k-1}, x_{i}^{k}\underline{\theta}_{i} \right] 
\left[
 \begin{array}{c}
U_{1}^{t} \\
V_{1}^{t}
 \end{array}
\right], \\
c_{i,k}
 &=
\left[ \underline{\omega}_{i},  x_{i}^{k-1}  \right] 
\left[
 \begin{array}{c}
 U_{1} \\
 U_{2}
 \end{array}
\right]  
+ x_{i}^{k}\underline{\theta}_{i} V_{2}^{t}, \\
\left[ c_{i,k+1},\dots,c_{i,\alpha} \right]
 &= 
\left[ \underline{\omega}_{i}, \ x_{i}^{k-1} \right] 
\left[
 \begin{array}{c}
 V_{1} \\
 V_{2}
 \end{array}
\right].
\end{align*} 
\item In the case of Type IV, the coding vector is represented as 
\begin{align*} 
\underline{\rho}_{i}
=
[1],  
\end{align*} 
where $k=1$ and $d=1$,  
and then, the components of the share are represented as
\begin{align*} 
c_{i,1}
&=
 U_{2}. 
\end{align*} 

\item In the case of Type V, the coding vector is represented as 
\begin{align*} 
\underline{\rho}_{i}
=
[1,
\underbrace{ x_{i},  \dots, x_{i}^{d-1} }_{x_{i}\underline{\theta}_{i}}
],  
\end{align*} 
where $k=1$ and $d\geq 2$, 
and then, the components of the share are represented as
\begin{align*} 
c_{i,1}
&=
 U_{2}
+ x_{i}\underline{\theta}_{i} V_{2}^{t}, \\
%
\left[ c_{i,2},\dots,c_{i,\alpha} \right]
&=
 V_{2}.
\end{align*} 
\end{enumerate}

In the case of Type I, 
since 
the message matrix is identical with that of the RSK-MSR code, 
the data collector can reconstruct the message from any $k$ shares 
by using the method of MSR Data-Reconstruction 
proposed by Rashmi et al.\cite[Theorem 5]{rashmi}. 

In the case of Type IV, 
the data collector can reconstruct the message from any share $\underline{c}_{i}$
without computing 
because $U_{2}=c_{i,1}$. 
Moreover, 
in the case of Type V, 
the data collector can reconstruct the message from any share $\underline{c}_{i}$ 
because 
$ V_{2}=\left[ c_{i,2},\dots,c_{i,\alpha} \right]$ 
and 
$U_{2}= c_{i,1} - x_{i}\underline{\theta}_{i} V_{2}^{t}$.

%
%


We show the reconstruction for Type III in the following theorem. 
The reconstruction for Type II is included in that for Type III. 
\begin{theorem}[Reconstruction for Type III]
\label{theorem:reconstructIII}
A data collector connecting any $k$ nodes can reconstruct the $B$ message symbols 
from the $k$ shares of the $k$ nodes.   
\begin{proof}
The data collector connects to $k$ nodes $\{ i_{1},  \dots, i_{k} \}$ to 
reconstruct the message, 
and downloads the $k$ shares $\underline{c}_{i_{1}},\dots,\underline{c}_{i_{k}}$. 

Decomposing k coding vectors 
$\underline{\rho}_{i_{1}},\dots,\underline{\rho}_{i_{k}}$ 
associated with the $k$ nodes, 
we define the following three matrices. 
\begin{align} 
\arraycolsep=2pt 
\Omega_{\text{DC}}
=
 \begin{bmatrix}
 \underline{\omega}_{i_{1}} \\
 \vdots \\
 \underline{\omega}_{i_{k}} 
 \end{bmatrix}, \ 
\underline{x}_{\text{DC}}
=
 \begin{bmatrix}
 x_{i_{1}}^{k-1} \\
 \vdots \\
 x_{i_{k}}^{k-1} 
 \end{bmatrix}, \ 
\Theta_{\text{DC}}
=
 \begin{bmatrix}
 x_{i_{1}}^{k}  \underline{\theta}_{i_{1}} \\
 \vdots \\
 x_{i_{k}}^{k}  \underline{\theta}_{i_{k}} 
 \end{bmatrix},
\end{align} 
where 
$\Omega_{\text{DC}}$ is a $(k \times (k-1))$ matrix, 
$\underline{x}_{\text{DC}}$ is a column vector of length $k$ and  
$\Theta_{\text{DC}}$ is a $(k \times (\alpha -k))$ matrix.  

\textit{(Step 1: Finding $V_{1}$ and $V_{2}$.)} 
Firstly, 
the data collector solves the following system of linear equations. 
\begin{align} 
\label{eq:step1}
 \begin{bmatrix}
 c_{i_{1},k+1} & \dots & c_{i_{1},\alpha}  \\
 \vdots & \vdots & \vdots \\
 c_{i_{k},k+1} & \dots & c_{i_{k},\alpha}  
 \end{bmatrix}
=
 \begin{bmatrix}
 \Omega_{\text{DC}} \  \underline{x}_{\text{DC}}
 \end{bmatrix}
 \begin{bmatrix}
 V_{1} \\
 V_{2}
 \end{bmatrix}.
\end{align} 
The left $(k \times k)$ matrix 
$\left[
\Omega_{\text{DC}} \ \underline{x}_{\text{DC}}
\right]$ 
in the right-hand side of Eq.(\ref{eq:step1}) 
is nonsingular,   
because the determinant of  
the $(k \times k)$ matrix is 
the Vandermonde determinant   
from the condition of $x_{i}$.  
Thus,  
the data collector can recover $ V_{1} $ and $ V_{2}$. 

\textit{(Step 2: Finding $U_{1}$ and $U_{2}$.)} 
Next, 
the data collector solves the following system of linear equations. 
\begin{align} 
\begin{bmatrix}
 c_{i_{1},k}  \\
 \vdots \\
 c_{i_{k},k}
\end{bmatrix}
-\Theta_{\text{DC}} V_{2}^{t}
=
\begin{bmatrix}
 \Omega_{\text{DC}} \ \underline{x}_{\text{DC}}
\end{bmatrix}
\begin{bmatrix}
 U_{1} \\
 U_{2}
\end{bmatrix}.
\end{align} 
Since 
the data collector knows $V_{2}$ from the previous step 
and 
the $(k \times k)$ matrix 
$\left[
\Omega_{\text{DC}} \ \underline{x}_{\text{DC}}
\right]$ 
is nonsingular, 
the data collector can recover $ U_{1} $ and $ U_{2}$. 
 
\textit{(Step 3: Finding $T_{1}$ and $T_{2}$.)} 
Finally,
the data collector recovers $T_{1}$ and $T_{2}$ 
in the right-hand side of Eq.(\ref{eq:step3})  
by using the method of MSR Data-Reconstruction\cite[Theorem 5]{rashmi}, 
that is, 
$T_{1}$ and $T_{2}$ are recovered by using the reconstruction method for Type I. 
\begin{align} 
\label{eq:step3}
\begin{bmatrix}
 c_{i_{1},1} & \dots & c_{i_{1},k-1}  \\
 \vdots & \vdots & \vdots \\
 c_{i_{k},1} & \dots & c_{i_{k},k-1}  
\end{bmatrix}
 -
\begin{bmatrix}
 \underline{x}_{\text{DC}} \ 
 \Theta_{\text{DC}}
\end{bmatrix}
\begin{bmatrix}
  U_{1}^{t} \\
  V_{1}^{t}
\end{bmatrix} 
=
\begin{bmatrix}
\Omega_{\text{DC}} \ 
\Lambda_{\text{DC}} \Omega_{\text{DC}}
\end{bmatrix}
\begin{bmatrix}
 T_{1} \\
 T_{2}
\end{bmatrix},
\end{align} 
where 
$\Lambda_{\text{DC}}$ 
is a $(k \times k)$ diagonal matrix with 
diagonal components $x_{i_{1}}^{\alpha},\dots,x_{i_{k}}^{\alpha}$. 
Note that 
the right-hand of Eq.(\ref{eq:step3}) is corresponding to \cite[Eq. (29)]{rashmi}.   
Since 
the data collector knows $U_{1}$ and $V_{1}$ from the previous steps, 
the data collector can obtain the values in the left-hand side of 
Eq.(\ref{eq:step3}). 
Thus, 
the data collector can recover $T_{1}$ and $T_{2}$ in the right-hand side of Eq.(\ref{eq:step3}) using the method of MSR Data-Reconstruction 
since  
the components in the right-hand side of Eq.(\ref{eq:step3}) satisfy 
the following conditions. 
\begin{itemize}
\item $\Omega_{\text{DC}}$ is $(k\times (k-1))$ matrix, 
\item $\Lambda_{\text{DC}}$ is a nonsingular $(k\times k)$ matrix from the condition of $x_{i}$, 
\item $T_{1}$ and $T_{2}$ are $((k-1)\times(k-1))$ symmetric matrices,  
\item 
the $((k-1)\times(k-1))$ matrix 
$
[
 \underline{\omega}_{i_{1}}^{t} \dots  \underline{\omega}_{i_{k-1}}^{t}
]^{t}
$
consisting of the first $k-1$ vectors 
$\underline{\omega}_{i_{1}}, \dots,  \underline{\omega}_{i_{k-1}}$ 
is nonsingular 
from the condition of $x_{i}$. 
\end{itemize}

By the above three steps, 
the data collector can recover the message matrix $M$, 
that is,  
the data collector can reconstruct the message consisting of $B$ message symbols. 
\end{proof}
\end{theorem}

From Theorem \ref{theorem:reconstructIII}, 
in the case of Type II, 
the data collector can reconstruct the message from any $k$ shares 
by the procedures of Step 2 and 3 
in the proof of Theorem \ref{theorem:reconstructIII}.

\subsection{Regeneration}
\label{subsec:regeneration}

In this section, 
we describe a regeneration of the generalized RSK-MSR code, 
that is, 
a failed node $f$ connecting any $d$ helper nodes can regenerate the share 
as was stored in itself prior to failure. 

Firstly, we describe the regeneration for Type I, II and III. 
For each type,   
we define the sub-vector $\underline{\varphi}_{i}$ of the coding vector $\underline{\rho}_{i}$ 
and the two sub-matrices $W_{1}$ and $W_{2}$ of the message matrix $M$, 
where 
$\underline{\varphi}_{i}$ is a vector of length $\alpha$,    
and 
$W_{1}$ and $W_{2}$ are respectively   
an $(\alpha \times \alpha)$ square matrix 
and 
a $((k-1) \times \alpha)$ matrix.  
\begin{enumerate}
\item 
In the case of Type I, 
the sub-vector $\underline{\varphi}_{i}$ and the two sub-matrices $W_{1}$, $W_{2}$ 
are defined as 
\begin{align*} 
\underline{\varphi}_{i}&=\underline{\omega}_{i} \in \mathbb{F}^{\alpha},\ 1 \leq i\leq n, \\
W_{1}&=T_{1},  W_{2}=T_{2}. 
\end{align*} 

\item 
In the case of Type II, 
the sub-vector $\underline{\varphi}_{i}$ and the two sub-matrices $W_{1}$, $W_{2}$ 
are defined as 
\begin{align*} 
\underline{\varphi}_{i}&=[\underline{\omega}_{i}, x_{i}^{k-1}] \in \mathbb{F}^{\alpha}, 
\ 1 \leq i\leq n, \\
W_{1}&=
\left[
\arraycolsep=2pt 
\begin{array}{cc}
T_{1}     & \multicolumn{1}{c}{U_{1}}  \\
U_{1}^{t} & \multicolumn{1}{c}{U_{2}}  
\end{array}
\right],  
W_{2}=
\left[
\arraycolsep=2pt 
\begin{array}{cc}
T_{2}  & O_{2}
\end{array}
\right].  
\end{align*}

\item 
In the case of Type III, 
the sub-vector $\underline{\varphi}_{i}$ and the two sub-matrices $W_{1}$, $W_{2}$ 
are defined as 
\begin{align*} 
\underline{\varphi}_{i}&=[\underline{\omega}_{i}, x_{i}^{k-1}, 
x_{i}^{k} \underline{\theta}_{i}
] \in \mathbb{F}^{\alpha}, \ 1 \leq i\leq n, \\
W_{1}&=
\left[
\arraycolsep=2pt 
\begin{array}{ccc}
T_{1}     & \multicolumn{1}{c}{U_{1}} & V_{1} \\
U_{1}^{t} & \multicolumn{1}{c}{U_{2}}     & V_{2} \\
V_{1}^{t} & \multicolumn{1}{c}{V_{2}^{t}} & O_{1} 
\end{array}
\right],  
W_{2}=
\left[
\arraycolsep=2pt 
\begin{array}{cc}
T_{2}  & O_{2}
\end{array}
\right].  
\end{align*} 
\end{enumerate}

From the above definitions, the share $\underline{c}_{f}$ of 
the failed node $f$ is represented as follows:
\begin{align} 
\underline{c}_{f}
=
 \underline{\varphi}_{f} W_{1} + x_{f}^{\alpha} \underline{\omega}_{f} W_{2}.
\end{align} 

The proof of the following thereom gives 
the regeneration method for Type I, II and III. 
The idea of the regeneration method is derived from 
the method of MSR Exact-Regeneration proposed by Rashmi, et al.\cite[Theorem 4]{rashmi}.   
\begin{theorem}[Regeneration for Type I, II and III]
\label{theorem:regeneration}
A failed node connecting any $d$ nodes can regenerate 
the share as was stored in itself prior to failure.   

\begin{proof}
The failed node $f$ connects $d$ helper nodes $\{ h_{1},\cdots,h_{d} \}$ 
to regenerate the same share that was stored in the failed node prior to failure, 
and downloads the following data for repair from each helper node. 
Each helper node $h_{p}$ computes data ${d}_{f,h_{p}}$ 
for repair from $\underline{c}_{h_{p}}$ and $\underline{\varphi}_{f}$ 
as follows: 
\begin{align} 
{d}_{f,h_{p}}
=
\underline{c}_{h_{p}} \underline{\varphi}_{f}^{t}
\in \mathbb{F}, 
\end{align} 
where 
${d}_{f,h_{p}}$ is a scalar because $\beta=1$, 
and sends it to the failed node $f$. 
As a result, the failed node obtains data of $d$ symbols 
$ {d}_{f,h_{1}} , \dots , {d}_{f,h_{d}}$  
for repair such that  
\begin{align} 
 \begin{bmatrix}
 {d}_{f,h_{1}} \\
 \vdots \\
 {d}_{f,h_{d}} 
 \end{bmatrix}
=
 \begin{bmatrix}
  \underline{\rho}_{h_{1}} \\
  \vdots \\
  \underline{\rho}_{h_{d}} 
 \end{bmatrix}
M
\underline{\varphi}_{f}^{t}
\in \mathbb{F}^{d} 
.
\end{align} 
Since the determinant of the leftmost $(d\times d)$ matrix in the right-hand side is 
the Vandermonde determinant from the condition of $x_{i}$,   
the $(d\times d)$ matrix is nonsingular. 
Thus, 
the failed node can obtain the following equation 
from the above system. 
\begin{align}
\label{eq:d-w1w2} 
 \begin{bmatrix}
  \underline{\rho}_{h_{1}} \\
  \vdots \\
  \underline{\rho}_{h_{d}} 
 \end{bmatrix}^{-1}
 \begin{bmatrix}
 {d}_{f,h_{1}} \\
 \vdots \\
 {d}_{f,h_{d}} 
 \end{bmatrix}
=
M
\underline{\varphi}_{f}^{t}
=
 \begin{bmatrix}
 W_{1} \underline{\varphi}_{f}^{t} \\
 W_{2} \underline{\varphi}_{f}^{t} 
 \end{bmatrix}
\end{align} 
Since the matrix $W_{1}$ is symmetric, 
we have $(W_{1} \underline{\varphi}_{f}^{t})^{t} = \underline{\varphi}_{f} W_{1}$. 
Since  $W_{2} \underline{\varphi}_{f}^{t} =  T_{2} \underline{\omega}_{f}^{t}$ and   
the matrix $T_{2}$ is symmetric, 
we have $(W_{2} \underline{\varphi}_{f}^{t})^{t} 
= (T_{2} \underline{\omega}_{f}^{t})^{t} 
= \underline{\omega}_{f} T_{2}$. 
Thus, the failed node can regenerate the share $\underline{c}_{f}$ as follows:
\begin{align} 
\label{eq:w1w2}
 (W_{1} \underline{\varphi}_{f}^{t}  )^{t} 
 +   x_{f}^{\alpha} \left[ (W_{2} \underline{\varphi}_{f}^{t} )^{t} , 
                           \underline{0}_{\alpha-k+1} \right]  
& = \underline{\varphi}_{f}W_{1}  +  
  x_{f}^{\alpha} \underline{\omega}_{f} \left[  T_{2} \  O_{2} \right] \\
& =
 \underline{\varphi}_{f}W_{1}  +  x_{f}^{\alpha} \underline{\omega}_{f} W_{2} 
=
\underline{c}_{f},
\end{align} 
where $\underline{0}_{\alpha-k+1}$ is a zero row vector of length $\alpha-k+1$ 
and 
$ \underline{\omega}_{f}  O_{2} = \underline{0}_{\alpha-k+1}$. 
\end{proof}
\end{theorem}
In the case of Type I, 
the regeneration method in the proof of Theorem \ref{theorem:regeneration} is 
identical with the method of MSR Exact-Regeneration\cite[Theorem 4]{rashmi}, 
where 
the zero vector $\underline{0}_{\alpha-k+1}$ 
and the zero matrix $O_{2}$ are removed from Eq.(\ref{eq:w1w2}) 
because $\alpha=k-1$. 

Next, we describe the regeneration for the remaining types IV and V.  
In the case of Type IV, 
the reconstruction is trivial because 
$\underline{c}_{i}=[c_{i,1}]=U_{2}$ and $d=1$. 
In the case of Type V, 
the failed node connecting any $d$ nodes can regenerate 
the share as was stored in itself prior to failure 
using the method of MBR Exact-Regeneration 
proposed by Rashmi et al.\cite[Theorem 2]{rashmi}    
because the message matrix $M$ is symmetric.

\section{Example}
\label{sec:example}

In this section, we give an example 
to help understanding of the generalized RSK-MSR code and 
the reconstruction and the regeneration of the code.  

Let $(n,k,d)=(10,2,4)$. 
Then $\alpha=d-k+1=3$ and $B=k\alpha=6$. 
From $(k,d)=(2,4)$, 
the message matrix $M$ corresponds to Type III. 
We can construct a $(10,2,4)$ generalized RSK-MSR code over $\mathbb{F}_{11}$  
from the following table for $\mathbb{F}_{11}$. 
\begin{align*}
\begin{array}{r | ccc ccc ccc c|}
\hline
i  & 
1 & 2 & 3 & 4 & 5 & 6 & 7 & 8 & 9 & 10 \\
\hline
\hline
\mathbb{F}_{11} \ni x_{i} & 
1 & 2 & 3 & 4 & 5 & 6 & 7 & 8 & 9 & 10 \\
\hline
x_{i}^{3}  & 
1 & 8 & 5 & 9 & 4 & 7 & 2 & 6 & 3 & 10 \\
\hline
\end{array}
\end{align*}
For each $i \in \{1,\dots,10\}$, 
we assign the element $x_{i} \in \mathbb{F}_{11} $ to node $i$.  

Let 
\begin{align*}
\arraycolsep=3pt 
M
=
\left[
\begin{array}{@{\hskip2pt} c|cc @{\hskip2pt} } 
T_{1} & \multicolumn{1}{c|}{U_{1}} & V_{1} \\
\hline
U_{1}^{t} & \multicolumn{1}{c|}{U_{2}} & V_{2} \\
\hline
V_{1}^{t} & \multicolumn{1}{c|}{V_{2}^{t}} & {O_{1}} \\
\hline
T_{2} & \multicolumn{2}{c}{O_{2}}
\end{array}
\right]
=
\left[
\begin{array}{@{\hskip2pt} c|cc @{\hskip2pt} } 
1 & \multicolumn{1}{c|}{2} & 3 \\
\hline
2 & \multicolumn{1}{c|}{4} & 5 \\
\hline
3 & \multicolumn{1}{c|}{5} & 0 \\
\hline
6 & 0 & 0 
\end{array}
\right],
\end{align*}
that is, the $B=6$ message symbols are $1,2,3,4,5$ and $6 \in \mathbb{F}_{11}$.  
For each $i \in \{1,\dots,10 \} $, 
the coding vector is given as 
$\underline{\rho}_{i}=[1,x_{i},x_{i}^{2},x_{i}^{3}]$, 
and then,  
the share $\underline{c}_{i}=\underline{\rho}_{i}M$ of node $i$ is given as 
\begin{align*}
\begin{bmatrix}
\underline{c}_{1} \\
\underline{c}_{2} \\
\underline{c}_{3} \\
\underline{c}_{4} \\
\underline{c}_{5} \\
\underline{c}_{6} \\
\underline{c}_{7} \\
\underline{c}_{8} \\
\underline{c}_{9} \\
\underline{c}_{10}
\end{bmatrix}
=
\arraycolsep=2pt 
\left[
\begin{array}{@{\hskip2pt} cccc @{\hskip2pt} } 
1 & 1 & 1 & 1 \\
1 & 2 & 4 & 8 \\
1 & 3 & 9 & 5 \\
1 & 4 & 5 & 9 \\
1 & 5 & 3 & 4 \\
1 & 6 & 3 & 7 \\
1 & 7 & 5 & 2 \\
1 & 8 & 9 & 6 \\
1 & 9 & 4 & 3 \\
1 &10 & 1 &10
\end{array}
\right]
M
=
\arraycolsep=2pt 
\left[
\begin{array}{@{\hskip2pt} cccc @{\hskip2pt} } 
1 & 0 & 8 \\
10& 8 & 2 \\
9 & 4 & 7 \\
1 &10 & 1 \\
0 & 4 & 6 \\
9 & 8 & 0 \\
9 & 0 & 5 \\
3 & 2 &10 \\
5 & 3 & 4 \\
7 & 3 & 9 
\end{array}
\right].
\end{align*}

\textit{(Reconstruction: )}
A data collector connects to $k=2$ nodes $\{ 1,2 \}$ to 
reconstruct the message, and 
downloads $k=2$ shares $\underline{c}_{1}$ and $\underline{c}_{2}$. 
Then, 
the matrices 
$\Omega_{\text{DC}}$, 
$\underline{x}_{\text{DC}}$ and 
$\Theta_{\text{DC}}$ are given as
\begin{align*} 
\Omega_{\text{DC}}
&=
\left[
 \begin{array}{@{\hskip2pt} c @{\hskip2pt} } 
 \underline{\omega}_{1} \\
 \underline{\omega}_{2} 
 \end{array}
\right]
=
\left[
 \begin{array}{@{\hskip2pt} c @{\hskip2pt} } 
1 \\
1
 \end{array}
\right], \ \ 
\underline{x}_{\text{DC}}
=
\left[
 \begin{array}{@{\hskip2pt} c @{\hskip2pt} } 
 x_{1} \\
 x_{2}
 \end{array}
\right]
=
\left[
 \begin{array}{@{\hskip2pt} c @{\hskip2pt} } 
1 \\
2
 \end{array}
\right], \ 
\Theta_{\text{DC}}
=
\left[
 \begin{array}{@{\hskip2pt} c @{\hskip2pt} } 
 x_{1}^{2} \underline{\theta}_{1} \\
 x_{2}^{2} \underline{\theta}_{2} 
 \end{array}
\right]
=
\left[
 \begin{array}{@{\hskip2pt} c @{\hskip2pt} } 
1 \\
4
 \end{array}
\right].
\end{align*} 

\textit{(Step 1: )} 
Firstly, 
the data collector recovers $V_{1}$ and $V_{2}$ as follows: 
\begin{align*} 
 \begin{bmatrix}
 V_{1} \\
 V_{2}
\end{bmatrix}
=
\begin{bmatrix}
  \Omega_{\text{DC}} \ \underline{x}_{\text{DC}}
\end{bmatrix}^{-1}
\begin{bmatrix}
 c_{1,3} \\
 c_{2,3}
\end{bmatrix} 
=
\begin{bmatrix}
 2 & 10  \\
 10& 1
\end{bmatrix} 
\begin{bmatrix}
 8  \\
 2
\end{bmatrix} 
=
\begin{bmatrix}
 3  \\
 5
\end{bmatrix}.
\end{align*} 

\textit{(Step 2: )} 
Next, 
the data collector recovers $U_{1}$ and $U_{2}$ as follows:
\begin{align*} 
\begin{bmatrix}
 U_{1} \\
 U_{2}
\end{bmatrix}
&=
\begin{bmatrix}
\Omega_{\text{DC}} \ \underline{x}_{\text{DC}}
\end{bmatrix}^{-1}
\left(
\begin{bmatrix}
 c_{1,2}  \\
 c_{2,2}
\end{bmatrix}
-\Theta_{\text{DC}} V_{2}^{t}
\right) \\
&=
\begin{bmatrix}
 2 & 10  \\
 10& 1
\end{bmatrix}
\left(
\begin{bmatrix}
 0  \\
 8
\end{bmatrix}
-
\begin{bmatrix}
 1  \\
 4
\end{bmatrix}
[5]
\right)
=
\begin{bmatrix}
 2  \\
 4
\end{bmatrix}.
\end{align*} 

\textit{(Step 3: )} 
Finally,
the data collector has the following system of linear equations 
from $U_{1}$ and $V_{1}$. 
\begin{align*} 
\label{eq:step3-example}  
\begin{bmatrix}
 \Omega_{\text{DC}} \ \Lambda_{\text{DC}} \Omega_{\text{DC}}
\end{bmatrix}
\begin{bmatrix}
 T_{1} \\
 T_{2}
\end{bmatrix}
&=
\begin{bmatrix}
 c_{1,1} \\
 c_{2,1}
\end{bmatrix}
 -
\begin{bmatrix}
 \underline{x}_{\text{DC}} \ 
 \Theta_{\text{DC}}
\end{bmatrix}
\begin{bmatrix}
  U_{1}^{t} \\
  V_{1}^{t}
\end{bmatrix}\\
&=
\begin{bmatrix}
 1 \\
 10
\end{bmatrix}
 -
\begin{bmatrix}
  1 & 1 \\
  2 & 4
\end{bmatrix}
\begin{bmatrix}
  2 \\
  3
\end{bmatrix}
=
\begin{bmatrix}
  7 \\
  5
\end{bmatrix}, 
\end{align*} 
where
\begin{align*} 
\Lambda_{\text{DC}} 
=
\begin{bmatrix}
      x_{1}^{3} & 0 \\
      0 & x_{2}^{3}
\end{bmatrix} 
=
\begin{bmatrix}
      1 & 0 \\
      0 & 8
\end{bmatrix}.
\end{align*} 
By using the method of MSR Data-Reconstruction\cite[Theorem 5]{rashmi},  
the data collector can recover $T_{1}$ and $T_{2}$ 
as follows:   
$T_{1}=[1]$ and $T_{2}=[6]$.     

From the above three steps, 
the data collector recovers the message matrix $M$, 
that is, 
the data collector reconstructs the $B=6$ message symbols. \\

\textit{(Regeneration: )}
Assume that node $1$ is a failed node, and 
node $2,3,4$ and $5$ are $d=4$ helper nodes for the failed node. 
The failed node obtains the following data of $d=4$ symbols for repair 
from the $d=4$ helper nodes. 
\begin{align*} 
\arraycolsep=2pt 
\begin{bmatrix}
 {d}_{1,2} \\
 {d}_{1,3} \\
 {d}_{1,4} \\
 {d}_{1,5} 
\end{bmatrix}
=
\begin{bmatrix}
 \underline{c}_{2} \\
 \underline{c}_{3} \\
 \underline{c}_{4} \\
 \underline{c}_{5}
\end{bmatrix}
 \underline{\varphi}_{1}^{t}
=
\begin{bmatrix}
 9 \\
 9 \\
 1 \\
 10 
\end{bmatrix}.
\end{align*} 
By computing 
the multiplication of two matrices 
in the left-hand side of  Eq.(\ref{eq:d-w1w2}), 
the failed node has the following equation.    
\begin{align*} 
\begin{bmatrix}
 W_{1} \underline{\varphi}_{1}^{t} \\
 W_{2} \underline{\varphi}_{1}^{t} 
\end{bmatrix}
&=
\begin{bmatrix}
\underline{\rho}_{2}  \\
\underline{\rho}_{3}  \\
\underline{\rho}_{4}  \\
\underline{\rho}_{5}  
%
\end{bmatrix}^{-1}
\begin{bmatrix}
 {d}_{1,2} \\
 {d}_{1,3} \\
 {d}_{1,4} \\
 {d}_{1,5}
\end{bmatrix} 
=
\begin{bmatrix}
10& 2 & 4 & 7 \\
5 & 8 & 1 & 8 \\
2 & 0 & 5 & 4 \\
9 & 6 & 5 & 2
\end{bmatrix} 
\begin{bmatrix}
9  \\
9  \\
1  \\
10
\end{bmatrix} 
=
\begin{bmatrix}
6 \\
0 \\
8 \\
6
\end{bmatrix} 
\end{align*} 
Next, 
the failed node 
substitutes 
the values such that 
$W_{1} \underline{\varphi}_{1}^{t} =[6,0,8]^{t}$ and 
$W_{2} \underline{\varphi}_{1}^{t}=[6]$ 
for 
Eq.(\ref{eq:w1w2}), and 
regenerate the share $\underline{c}_{1}$ as follows: 
\begin{align*} 
\underline{c}_{1}
&= (W_{1} \underline{\varphi}_{1}^{t})^{t} +  x_{1}^{3} 
\left[ (W_{2} \underline{\varphi}_{1}^{t})^{t} , \underline{0}_{2} \right] \\
&= [6,0,8] +  [6,0,0] = [1,0,8],
\end{align*} 
where 
$\underline{0}_{2} = [0,0]$.

\section{Relations Between $\{\ell,\ell'\}$ Secure Codes and Generalized RSK-MSR Codes}
\label{sec:securecode}

In this section, 
we first describe the concept of the $\{\ell,\ell'\}$ secure code and 
the construction method proposed by Shah, Rashmi and Kumar\cite{shah2}. 
Next, 
we show how to apply 
the construction method of $\{\ell,\ell'\}$ secure code 
to 
the generalized RSK-MSR code.   
Finally, 
we explain that 
the message matrix of the generalized RSK-MSR code 
is derived from that of the original RSK-MSR code 
by using the construction method of the $\{\ell=k,\ell'=0\}$ secure code.

\subsection{$\{\ell,\ell'\}$ Secure Codes based on Product-Matrix MSR Codes}

Shah, Rashmi and Kumar\cite{shah2} 
proposed the construction methods of 
the $\{\ell,\ell'\}$ secure codes based on 
the $(n,k,d)$ Product-Matrix(PM) MSR codes and the $(n,k,d)$ PM MBR codes, 
which are proposed by 
Rashmi et al.\cite{rashmi}.  
Note that 
the $(n,k,d)$ PM MSR code is identical with the $(n,k,d)$ RSK-MSR codes with 
the $(2\alpha \times \alpha)$ message matrix in the case of Type I. 

In \cite{shah2}, 
the $\{\ell,\ell'\}$ secure codes are defined in the following threat model. 
An eavesdropper can gain read-access to the data stored 
in any set of at-most $\ell(<k)$ storage nodes. 
The eavesdropper may also gain read-access to the data being downloaded 
during (possibly multiple instances of) repair of some $\ell'(\leq \ell)$ 
of these $\ell$ nodes. 
Note that the data downloaded by a replacement node during any instance of repair also 
contains the data that is eventually stored in that node.   
This is formalized in the following definition\cite[Definition 1]{shah2}.
\begin{definition}[$\{\ell,\ell'\}$ secure distributed storage system\cite{shah2}]
Consider a distributed storage system in which an eavesdropper gains access to 
the data stored in some $(\ell-\ell')$ nodes, 
and the data stored as well as the data downloaded during repair in some other $\ell'$ nodes. 
An $\{\ell,\ell'\}$ secure distributed storage system is one in which 
such an eavesdropper obtains no information about the message. 
\end{definition}

We describe the construction method of the $\{\ell,\ell'\}$ secure code\cite[Sec.  IV-B]{shah2}.  
The form of the message matrix of the $(n,k,d)$ PM MSR code is a $(2\alpha\times\alpha)$ 
matrix 
$
\left[ 
\begin{smallmatrix}
S_{1}\\
S_{2}
\end{smallmatrix}
\right]
$, 
where 
$S_{1}$ and $S_{2}$ are $(\alpha\times\alpha)$ symmetric matrices. 
To construct an $\{\ell,\ell'\}$ secure code based on the $(n,k,d=2k-2)$ PM MSR code 
with $(\alpha=k-1, \beta=1, B=k\alpha)$ over $\mathbb{F}_{q}$,  
in the input to the $(n,k,2k-2)$ PM MSR code (without secrecy), 
a specific, carefully chosen set of $R=\ell \alpha+(k-\ell)\ell'$ message symbols 
in the $(2\alpha \times \alpha)$ message matrix   
are replaced with 
$R$ random symbols as follows.  
Each of these random symbols are chosen uniformly and independently from 
$\mathbb{F}_{q}$, and are also independent of the message symbols. 
Use these $R$ random symbols to replace the following $R$ symbols 
in the $(2\alpha \times \alpha)$ message matrix 
$
\left[ 
\begin{smallmatrix}
S_{1}\\
S_{2}
\end{smallmatrix}
\right]
$, 
to obtain 
matrix $M^{(s)}$ 
of the $\{\ell,\ell'\}$ secure code: 
\begin{itemize}
\item
the $\ell \alpha- \tbinom{\ell}{2}$ symbols in 
the first $\ell$ rows 
(and hence the first $\ell$ columns) of the $(\alpha \times \alpha)$ symmetric matrix $S_{1}$, 
\item
the $\binom{\ell}{2}$ symbols in 
the intersection of 
the first $(\ell-1)$ rows 
and 
the first $(\ell-1)$ columns of the $(\alpha\times\alpha)$ symmetric matrix $S_{2}$, 
\item 
the $(k-\ell)\ell'$ remaining symbols in the first $\ell'$ rows 
(and hence the first $\ell'$ columns) of $S_{2}$. 
\end{itemize}
For each $i$, $1 \leq i \leq n$, 
the data stored in node $i$, i.e., the share $\underline{c}_{i}$ of node $i$, 
is given by 
$\underline{c}_{i}=[1,x_{i},x_{i}^{2},\cdots,x_{i}^{2\alpha-1}]M^{(s)}$, 
where 
$[1,x_{i},x_{i}^{2},\cdots,x_{i}^{2\alpha-1}]$ is identical with 
the coding vector $\underline{\rho}_{i}$ in the case of Type I.

The $\{\ell,\ell'\}$ secure codes guarantee the following secrecy.  
Let $\mathcal{U}$ denote the collection of the $(B-R)$ message symbols, 
and 
let $\mathcal{R}$ denote the collection of $R$ random symbols.  
Further, 
let $\mathcal{E}$ denote the collection of symbols 
that an eavesdropper gains access to. 
And then,  
the eavesdropper obtains no information about the message from 
the data stored in some $(\ell-\ell')$ nodes  
and the data stored as well as the data downloaded during repair in some other $\ell'$ nodes, 
that is,  
the mutual information between the message symbols $\mathcal{U}$ 
and 
the symbols $\mathcal{E}$ obtained by the eavesdropper is zero, 
i.e., 
$I(\mathcal{U};\mathcal{E})=0$,   
where 
$I(\mathcal{U};\mathcal{E})$ denotes a mutual information between $\mathcal{U}$ and $\mathcal{E}$, 
and  
all logarithms are taken to the base $q$.

\subsection{$\{\ell,\ell'\}$ Secure Codes based on Generalized RSK-MSR Codes}

In the cases of Type II and III,    
we show how to apply 
the construction method of $\{\ell,\ell'\}$ secure code\cite[Sec.  IV-B]{shah2} 
to the $(n,k,d)$ generalized RSK-MSR code  
with the $(d \times \alpha)$ message matrix $M$.  
To construct an $\{\ell,\ell'\}$ secure code based on 
the $(n,k,d)$ generalized RSK-MSR code, which has the message matrix of Type II and III, 
we use $R=\ell \alpha + (k-\ell)\ell'$ random symbols and 
$(B-R)$ message symbols. 
Use these $R$ random symbols to replace the following $R$ symbols 
in the $(d\times \alpha)$ message matrix $M$, 
to obtain matrix $M^{(s)}$ 
of the $\{\ell,\ell'\}$ secure code:  
\begin{itemize}
\item 
the $\ell \alpha- \tbinom{\ell}{2}$ symbols 
in the first $\ell$ rows (and hence the first $\ell$ columns) of 
the $(\alpha \times \alpha)$ symmetric matrix $W_{1}$, 
\item 
the $\tbinom{\ell}{2}$ symbols in 
the intersection of 
the first $(\ell-1)$ rows 
and 
the first $(\ell-1)$ columns 
of the $((k-1)\times(k-1))$symmetric matrix $T_{2}$, 
\item 
the $(k-\ell)\ell'$ remaining symbols 
in the first $\ell'$ rows (and hence the first $\ell'$ columns) of $T_{2}$. 
\end{itemize}
For each $i$, $1 \leq  i \leq n$, 
the share $\underline{c}_{i}$ stored in node $i$ is given by 
$\underline{c}_{i}=\underline{\rho}_{i}M^{(s)}$.

\subsection{
Relation between message matrices of 
$\{\ell=k,\ell'=0\}$ secure code and $(n,k,d)$ Generalized RSK-MSR Code} 

We explain that 
the $(d\times \alpha)$ message matrix $M$ of 
an $(n,k,d)$ generalized RSK-MSR code 
is derived from that of the PM MSR code 
by using the construction method of an $\{\ell=k,\ell'=0\}$ secure code.  

To construct the $(d \times \alpha)$ message matrix $M$ of Type II, III, IV and V,  
we prepare the $(2 \alpha \times \alpha)$ message matrix 
$
\left[ 
\begin{smallmatrix}
S_{1}\\
S_{2}
\end{smallmatrix}
\right]
$ 
of  
an $(\widehat{n}, \widehat{k},\widehat{d})$ PM MSR code 
with $(\widehat{\alpha}, \widehat{\beta}=1,\widehat{B})$ 
such that 
$\widehat{d} = 2 \widehat{k}-2$, 
where 
$\widehat{k}=d-k+2$. 
From the condition $\widehat{k}=d-k+2$, 
it holds that $\widehat{k}=\alpha+1$, $\widehat{d}=2\alpha$ and 
$\widehat{B}=\widehat{k}\widehat{\alpha}=\alpha(\alpha+1)\geq B$  
because $\alpha=d-k+1$ and $d \geq 2k-2$, 
where $\alpha=d-k+1$ and $B=k\alpha$.   
The $(\widehat{n}, \widehat{k},\widehat{d})$ PM MSR code 
is the underlying code of the  $\{\ell=k,\ell'=0\}$ secure code.    

Firstly, 
by using the construction method of $\{\ell=k,\ell'=0\}$ secure code, 
we replace 
the following $B$ symbols 
in 
the $(2\alpha\times\alpha)$ matrix 
$
\left[ 
\begin{smallmatrix}
S_{1}\\
S_{2}
\end{smallmatrix}
\right]
$ 
with 
$B$ message symbols as follows:   
\begin{enumerate}
\item In the cases of Type II and III (i.e., $2 \leq k \leq \alpha$), 
\begin{itemize}
\item 
$B-\tbinom{k}{2}$ symbols in 
the first $k$ rows of the $(\alpha \times \alpha)$ symmetric matrix $S_{1}$, 
\item 
the $\tbinom{k}{2}$ symbols in 
the intersection of 
the first $(k-1)$ rows 
and 
the first $(k-1)$ columns 
of the $(\alpha\times\alpha)$ symmetric matrix $S_{2}$. 
\end{itemize}

\item 
In the cases of Type IV and V (i.e., $k =1$),  
$B$ symbols in 
the first row of the $(\alpha \times \alpha)$ symmetric matrix $S_{1}$, 
where $B=d=\alpha$.  
\end{enumerate}

Next, 
we replace the remaining $(\widehat{B}-B)$ symbols in 
the $(2\alpha\times\alpha)$ matrix 
$
\left[ 
\begin{smallmatrix}
S_{1}\\
S_{2}
\end{smallmatrix}
\right]
$ 
with 
$(\widehat{B}-B)$ zeros.

As a result, 
we obtain the matrix 
$
\left[ 
\begin{smallmatrix}
S_{1}\\
S_{2}
\end{smallmatrix}
\right]
$ 
such that 
the $((\alpha-k+1)\times\alpha)$ lower sub-matrix of 
$
\left[ 
\begin{smallmatrix}
S_{1}\\
S_{2}
\end{smallmatrix}
\right]
$ 
is a zero matrix 
and  
the $(d \times\alpha)$ upper sub-matrix of  
$
\left[ 
\begin{smallmatrix}
S_{1}\\
S_{2}
\end{smallmatrix}
\right]
$ 
is identical with the 
$(d\times \alpha)$ 
message matrix $M$ of 
the $(n,k,d)$ generalized RSK-MSR code.

\section{Conclusion}
\label{sec:conclusion}

We have proposed 
the construction of $(n,k,d)$ MSR codes 
for arbitrary $(n,k,d)$ such that $d \geq 2k-2$ and $d \leq n-1$. 
The proposed MSR code is 
the generalized version of 
the Rashmi-Shah-Kumar Minimum-Storage-Regenerating(RSK-MSR) code 
based on the product-matrix framework. 
In addition, 
we have described 
the relations between the generalized RSK-MSR codes and $\{\ell,\ell'\}$ secure codes.


%



%
%

\ifCLASSOPTIONcaptionsoff
  \newpage
\fi


\begin{thebibliography}{9}

\bibitem{dimakis}
A. G. Dimakis, P. B. Godfrey, Y. Wu, M. J. Wainwright and K. Ramchandran,  
``Network Coding for Distributed Storage Systems,''
IEEE Trans. Inf. Theory, 
vol. 56, no. 9, pp. 4539-4551, Sept. 2010.

\bibitem{rashmi}
K. V. Rashmi, N. B. Shah, and P. V. Kumar, 
``Optimal Exact-Regenerating Codes for Distributed Storage at 
the MSR and MBR Points via a Product-Matrix Construction,'' 
IEEE Trans. Inf. Theory, vol. 57, no. 8, 
pp.5227-5239, Aug. 2011.

\bibitem{suh}
C. Suh and K. Ramchandran, 
``Exact-repair MDS code construction
using interference alignment,'' 
IEEE Trans. Inf. Theory, vol. 57, no. 3,
pp. 1425-1442, Mar. 2011.

\bibitem{shah}
N. B. Shah, K. V. Rashmi, P. V. Kumar, and K. Ramchandran,
``Interference Alignment in Regenerating Codes 
for Distributed Storage: Necessity and Code Constructions,''
IEEE Trans. Inf. Theory, vol. 58, no. 4,
pp.2134-2158, Apr. 2012. 

\bibitem{shah2}
N.B.Shah,  K.V.Rashmi, and P.V.Kumar, 
 ``Information-Theoretically Secure Regenerating Codes 
for Distributed Storage,'' 
{in 
IEEE Global Communications Conference (GLOBECOM) 2011, Houston, Texas, USA, Dec. 2011. }

\end{thebibliography}
\end{document}